\documentclass[usenatbib]{mnras}
\usepackage{newtxtext,newtxmath}
\usepackage[T1]{fontenc}

\DeclareRobustCommand{\VAN}[3]{#2}
\let\VANthebibliography\thebibliography
\def\thebibliography{\DeclareRobustCommand{\VAN}[3]{##3}\VANthebibliography}

\usepackage{amsmath}
\usepackage{graphicx}
\usepackage[dvipsnames]{xcolor}
\usepackage{mathtools}
\usepackage{gensymb}
\usepackage{tabularx}
\usepackage{subcaption}
\usepackage{physics}
\usepackage{subfiles}
\usepackage{multicol}
\usepackage[export]{adjustbox}

\newcommand{\dobib}{\bibliography{references}}
\newcommand{\ccm}{\,\ifmmode{\mathrm{cm}^{-3}}\else cm$^{-3}$ \fi}
\title[Neural Infalling Cloud Equations (NICE)]{Neural Infalling Cloud Equations (NICE): Increasing the Efficacy of Subgrid Models and Scientific Equation Discovery using Neural ODEs and Symbolic Regression}

\author[B. Tan] {
  \href{https://orcid.org/0000-0003-4805-6807}{Brent Tan}$^1$\thanks{E-mail: btan@flatironinstitute.org}\\
    $^1$Center for Computational Astrophysics, Flatiron Institute, 162 5th Ave, New York, NY 10010, USA
}
\date{Accepted XXX. Received YYY; in original form ZZZ}
\pubyear{2024}

\begin{document}
\renewcommand{\dobib}{}
\label{firstpage}
\pagerange{\pageref{firstpage}--\pageref{lastpage}}
\maketitle

\begin{abstract}
Galactic systems are inherently multiphase, and understanding the roles and interactions of the various phases is key towards a more complete picture of galaxy formation and evolution. For instance, these interactions play a pivotal role in the cycling of baryons which fuels star formation. The transport and dynamics of cold clouds in their surrounding hot environment are governed by complex small scale processes (such as the interplay of turbulence and radiative cooling) that determine how the phases exchange mass, momentum and energy. Large scale models thus require subgrid prescriptions in the form of models validated on small scale simulations, which often take the form of coupled differential equations. 
In this work, we explore using neural ordinary differential equations which embed neural networks as terms in the model to capture an uncertain physical process. We then apply Symbolic Regression to potentially discover new insights into the physics of cloud-environment interactions. We test this on both generated mock data and actual simulation data. We also extend the neural ODE to include a secondary neural term. We show that neural ODEs in tandem with Symbolic Regression can be used to enhance the accuracy and efficiency of subgrid models, and/or discover the underlying equations to improve generality and scientific understanding. We highlight the potential of this scientific machine learning approach as a natural extension to the traditional modelling paradigm, both for the development of semi-analytic models and for physically interpretable equation discovery in complex non-linear systems.
\end{abstract}

\begin{keywords}
% choose six from https://academic.oup.com/DocumentLibrary/mnras/keywords.pdf
methods: data analysis -- hydrodynamics -- instabilities -- turbulence -- galaxies: haloes -- galaxies: evolution
\end{keywords}

%%%%%%%%%%%%%%%%%%%%%%%%%%%%%%%%%%%%%%%%%%%%%%%%%%%%%%%%%%%%%%%%%%%%%%%%%%%%%%%%%%
%                                   Body                                         %
%%%%%%%%%%%%%%%%%%%%%%%%%%%%%%%%%%%%%%%%%%%%%%%%%%%%%%%%%%%%%%%%%%%%%%%%%%%%%%%%%%

\section{Introduction}
A key open question in galaxy evolution is the discrepancy between observed star formation rates in galaxies and their available cold (T $\sim 10^4$\,K) gas reservoirs. The observed star formation rates are unsustainable over cosmic timescales, indicating the need for continuous cold gas accretion to fuel star formation \citep{erb08,hopkins08,putman09}. For example, our Milky Way would burn through its current supply in a mere 2 Gyrs \citep{chomiuk11,putman12}! This necessitates a detailed understanding of how cold gas is transported between different regions of galaxies. 

There is plenty of evidence for the existence of the gas flows which mediate this transport. Widespread observations find cold material outflowing at super-virial/escape velocities \citep{veilleux05,steidel10,rubin14,heckman17}. We also observe infall in the form of  ‘high-velocity’ and ‘intermediate-velocity’ clouds (HVCs and IVCs; \citealt{putman12}) and see evidence of fountain-like transport \citep{shapiro76, fraternali08}, where cold clouds are lifted out of the disk by supernova-driven winds and then fall back in. In many ways, these processes are analogous to the water cycle, but on galactic scales, and are hence known as the Galactic Baryon Cycle. Feedback mechanisms (either from stars or active galactic nuclei) drive Galactic outflows which carry material outwards into the circumgalactic medium (CGM) \citep{veilleux05}, while inflowing gas fuels new star formation \citep{ keres05, dekel06, fraternali15}. This cycling hence links processes on stellar ($\sim$\,pc) scales to galactic ($\sim$\,kpc) scales.

One of the primary challenges in modeling these flows is accurately simulating the complex interactions between the cold gas clouds and their hot (T $\sim 10^6$\,K) ambient medium, which is sensitive to even smaller sub-pc scale processes at their interfaces. For example, hydrodynamic instabilities such as the Kelvin-Helmholtz instability (KHI) pose significant difficulties for survival as they can quickly shred the clouds into smaller fragments, potentially leading to their complete destruction. The cloud's ability to survive is also influenced by various other factors \citep{gronke18}, including its size, density, and the efficiency of radiative cooling (on a cooling timescale $t_{\rm cool} \lesssim 1$\,Myr) within the turbulent mixing layers formed at the interface between the cold and hot gases \citep{begelman90,fielding20,tan21}. Accurately capturing the fast cooling small-scale radiative turbulent mixing layers in simulations is hence crucial but challenging due to the need for high spatial (sub-pc) and temporal ($\ll t_{\rm cool}$) resolution, which are not presently reachable even for high resolution galaxy scale simulations \citep[e.g.,][]{schneider20, steinwandel22, rey23} (which can resolve scales of order 5\,pc), and even less so for the subset of large scale cosmological simulations which are able to self consistently develop multiphase structures (on scales of order 100\,pc) in the CGM \citep{nelson20,ramesh23}.

One way around this is to develop subgrid models that capture the effects of unresolved processes on the larger resolved scales. These models are essential for simulating the evolution of galactic clouds in the larger scale simulations described above, where the the range of scales renders it computationally untenable to resolve the detailed structure of the clouds and their interfaces. Such models can be formulated and tested with smaller scale high resolution simulations and then included in the larger scale simulations directly \citep[e.g.,][for clouds]{tan23, tan24, smith24}, or in semi-analytical models \citep[e.g.,][for galactic winds]{fielding22}. Improving the fidelity of these subgrid models is thus important for advancing galaxy simulations. This can be done in two ways: better calibrating the models to simulations that resolve the relevant length scales, or improving our understanding of the underlying physics so that these models generalize well.

In this spirit, \citet{tan23} recently developed a simple analytic theory and model for infalling clouds based on the physics of turbulent radiative mixing layers. They formulate a subgrid model for the evolution of these clouds, comprising of a set of couple ordinary differential equations (ODEs). While their model generally captures the evolution of the clouds well, it makes certain assumptions, such as an ansatz for a weight factor to account for the growth of turbulent mixing layers over some initial development time. While this process can be seen visually in the actual simulations, forming this ansatz required guessing a functional form and tuning it to match the simulation data. This is thus the least physically motivated part of the model. In this work, we explore using neural ODEs to learn the evolution of the weight factor directly from the data instead. This will allow us to improve the predictive accuracy of the subgrid model. We also explore the use of Symbolic Regression (SR) on the learned model to potentially discover new insights into the physics of cloud-environment interactions.

Neural ODEs are ODEs which include neural networks in their parameterizations of the vector field and differ from traditional deep neural network architectures in that they replace hidden layers with the continuous transformation governed by the ODE \cite{chen18}. One way to see this is that there is hence a traditional neural network embedded in a differential equation, which in turn is embedded in a non-traditional neural network encompassing the overall computation graph that feeds through the ODE solver. They can also be seen as a special case of the more general class of Universal Differential Equations which can include any universal approximator \cite{rackauckas20}. Because of their structure, neural ODEs naturally inherit both the advantages and disadvantages of having to work with an ODE solver. For example, it is easy to balance numerical precision and speed by adjusting the timestep (and hence depth) of integration of the ODE for simple or smooth systems, but stiff ODEs on the other hand can be challenging to solve and the choice of a solver introduces yet more hyperparameters.
Neural ODEs have been applied in a range of domains, including generative modelling, physical modelling, and time series modelling. Examples include neuroscience \cite{kim21}, fluid mechanics \cite{brunton20}, climate science \cite{hwang21} and epidemiology \cite{kuwahara23}, just to name a handful.

This is a rich topic of ongoing research (see \citet{kidger22} for a recent survey), but a brief overview is as follows. The core idea is that we can treat neural ODEs as differentiable systems, including the ODE solver. We can hence perform backpropogation and thus train via the usual methods such as stochastic gradient descent. Alternatively, neural ODEs can be treated as the natural continuous limit of many modern deep learning architectures such as residual and recurrent networks. In other words, such architectures naturally embed the ability to characterize high dimensional vector fields directly into their structure. Neural ODEs are distinctly different from Physics-informed neural networks (PINNs), which do not directly include the differential equations in their architecture but instead incorporate them into the cost functions of models so as to incentivize certain priors (i.e., the physics). PINNs have already seen rapid adoption in astronomy for a variety of applications such as obtaining quasinormal modes of non-rotating black holes \citep{cornell22}, creating fast emulators for chemical solvers in hydrodynamic simulations \citep{branca23}, and generating hydrodynamic information for dark matter only simulations \citep{zhenyu24}. In neural ODEs, the structure imposed by differential equation essentially imposes these strong physical priors on the model space. Theory-driven differential equation models are already the classic workhorse of scientific modelling in the physical sciences. However, these models are limited by the ability of the theory to capture the details of the actual processes, and all the assumptions and approximation contained wherein. There is almost always some residual difference between the data and the model that arises from this incompleteness. Neural ODEs offer a way to combine these traditional models with deep learning and its strength in serving as high-capacity function approximators \cite{lecun15} to learn the residual difference directly from the data, and hence improve the predictive power of the model. By restricting the scope of the neural network in the differential equations, we can also ensure that the model remains interpretable and physically motivated. This is in contrast to the usual image of deep neural networks as black boxes which are often criticized for their lack of interpretability (we note that there exists various methods which aim to address this issue, see for example Shapley Additive Explanations \citep{lundberg17} which uses Shapley values from game theory to evaluate the importance of each input feature on the prediction of a neural network). Neural ODEs attempt to combine the best of both worlds. 
They have thus been recently highlighted as a promising discovery tool with mechanistic interpretability on both mock and real observations of galactic winds \citep{nguyen23a, nguyen23b}. 

This opens the door to using Symbolic Regression effectively to discover new insights into the underlying physics of the system \citep{cranmerDiscovering2020}. Symbolic Regression is a type of regression technique that models a given dataset by constructing analytic functions, typically searching equation space through the use of genetic algorithms. Starting with a random population of mathematical expressions and a metric to evaluate the fitness of each expression (used to score how well they fit the data), the population can be iteratively refined via mutation (random changes), crossovers (hybridizing equations) and selection (keeping the best equations) to keep refining the equations. This sets it apart from traditional regressions techniques such as linear regression which assume a predefined model structure, making it more flexible. Consequentially, one primary challenge of Symbolic Regression is how quickly the search space scales with complexity of the dataset (essentially exponentially with the number of input variables and allowed operators). Hence, by limiting the scope of the physical process the neural network is responsible for capturing, we can increase the computational tractability of applying Symbolic Regression methods to the trained embedded neural networks in the neural ODEs.

We will show that neural ODEs in tandem with Symbolic Regression can be used to enhance the accuracy and efficiency of subgrid models, and/or discover the underlying equations to improve generality and scientific understanding. While this work stays within the context of galactic cloud-environment interactions, we believe this naturally integrates seamlessly, flexibly and rapidly into existing traditional scientific modelling workflows, and should hence be included and leveraged as an essential part of any modern modelling toolkit. The outline of this paper is as follows. In Section~\ref{sect:methods}, we provide a description of the physical problem and the model, the setup of the numerical simulations, and the architecture and training strategy employed in this work. We present our results in Section~\ref{sect:results}, and discuss some limitations before summarizing and concluding in Section~\ref{sect:conclusions}.

\section{Methods} \label{sect:methods}
\subsection{Background: Infalling Cold Clouds}
We begin by describing the physical problem at hand. As mentioned in the introduction, the cycling of baryons, and in particular the cold gas which fuels star formation, plays an important role in galaxy formation. This cycle includes both outflows from and inflows into the galactic disc, both of which are multiphase in that they involve both cold gas clouds and background hot gas. Both also involve velocity shears between the two temperature phases, which was the source of a long-standing problem regarding the survival of the cold gas clouds since they should be rapidly destroyed by hydrodynamic instabilities (in particular the Kelvin-Helmholtz instability). For a cloud of radius $r$ with overdensity $\chi$ relative to the background and embedded in a wind of velocity $v_{\rm w}$, the entrainment time $t_{\rm acc}$ is roughly the timescale for the cloud to sweep up its own column density $\sim \chi r/v_{\rm w}$. However, the destruction timescale via instabilities, also know as the `cloud crushing' time, is $t_{\rm cc} \sim \sqrt{\chi} r/v_{\rm w}$. Since $t_{\rm acc}/t_{\rm cc} \sim \sqrt{\chi}$, it was expected that clouds should be destroyed before they can be accelerated and entrained \citep{klein94,zhang17}. This theoretical argument held up in a host of simulation studies (including those with radiative cooling) \citep[e.g.][]{cooper09,scannapieco15,schneider16}. The inclusion of magnetic fields did not qualitatively change this picture \citep{mccourt15,gronke20-cloud}.

However, it was recently found that if the cooling efficiency of mixed `warm' gas at the interfaces of the phases was large enough, and if mixing was enhanced by turbulence, then there could be a significant amount of transfer of mass, momentum and energy from the hot to the cold phase \citep{begelman90,ji18,fielding20,tan21}. The criteria for this to happen is $t_{\mathrm{cool, mix}}/t_{\mathrm{cc}} < 1$, where $t_{\mathrm{cool,mix}}$ is the cooling time of the mixed warm gas (with $T_{\rm mix} \sim (T_{\rm hot} T_{\rm cold})^{1/2}$) \citep{gronke18}. For clouds, this is satisfied when the cloud is large enough such that $r > c_{\rm s, cold} t_{\rm cool,mix}$ (where $c_{\rm s, cold}$ is the sound speed of the cold gas). In this regime, the cloud grows by accreting material and hence momentum from the background which speeds up the timescale for entrainment. Critically, it was also found that the box size of the simulation needed to be long enough to capture the growth of a `tail' behind the head of the cloud where much of the growth eventually happens \citep{gronke18}. These conclusions have since held up in many subsequent numerical studies \citep[e.g.,][]{sparre20,li20,abruzzo21,girichidis2021,farber22}.

Most of the recent work in this area has focused on the outflowing part of the baryon cycle, but inflows of clouds are arguably just as important for fueling ongoing star formation. \citet{tan23} thus focused on the survival and growth of cold gas clouds that are instead \textit{infalling} towards the galactic disc under gravity. Observationally, these are often observed as high velocity clouds (HVCs). HVCs were first detected in HI\,21\,cm emission \citep{muller63} and are defined as gas clouds observed to be moving at high velocities relative to the local standard of rest ($|v_{\rm LSR}| \geq 90$\,km/s) \citep{wakker91}. They have been observed at a range of sky locations and sizes, and could provide a large amount of the necessary fuel for star formation if they survive their journey to the disc \citep{vanwoerden,putman12,fox19}. Initial efforts to model HVCs focused on ballistic velocity  trajectories \citep{bregman80} with some estimates of hydrodynamic drag \citep{benjamin&daly97}, and these models were then used to estimate HVC contributions in feedback models \citep{maller04}. With the advent of the high resolution simulations which show the coupling of velocity and mass evolution of the clouds to the background they move through in winds, these HVC models are also starting to be updated \citep{heitsch09,heitsch22,gronnow22,tan23}. Unsurprisingly, the survival and growth of HVC-like clouds are quite different from the wind tunnel clouds, primarily due to gravitational acceleration. Below, we review the model for these infalling clouds developed in \citep{tan23}, which they tested against a suite of simulations.

As a last background note, modelling HVCs is also important for empirical tests of the complex physics at play in the radiative turbulent mixing layers. Most observational data of clouds in outflowing winds are at extra-galactic distances and hence lacking in resolution. In contrast, there is a large body of much better resolved (both spatially and kinematically) observational data for HVCs in the Milky Way. They are hence the best test bed we have for constraining theories of processes such as multiphase turbulent mixing and radiative cooling.

\subsection{Theoretical Model}
In this section, we summarize the model developed in \citep{tan23}. We can describe the evolution of an infalling cloud growing via mixing-cooling driven accretion of the background CGM by the following set of coupled ODES:
\begin{align} 
    \frac{\dd z}{\dd t}    &= v \label{eq:diff_eq1} \\
    \frac{\dd (mv)}{\dd t} &= mg - \frac{1}{2}\rho_{\rm hot} v^2 C_0 A_{\rm cross} \label{eq:diff_eq2} \\
    \frac{\dd m}{\dd t}    &= \frac{m}{t_{\rm grow}(z,v,m)} 
    \label{eq:diff_eq3}
\end{align}
where $z$, $v$ and $m$ represent the distance fallen, the velocity and the mass of the cloud respectively. $t_{\rm grow} \equiv m/\Dot{m}$ is the growth timescale, $g$ is the gravitational acceleration, $C_0$ is a geometry dependent drag coefficient of order unity, $\rho_{\rm hot}$ is the density of the hot background medium, and $A_{\rm cross}$ is the cross-sectional area the cloud presents to the background flow. An important assumption in this model is that equation~\eqref{eq:diff_eq3} assumes continuous growth, omitting terms which might contribute to cloud destruction. This model hence only applies to clouds which are gaining rather than losing mass. This is valid when the growth timescale $t_{\rm grow}$ is shorter than the destruction timescale $t_{\rm cc}$. This model only considers the hydrodynamic case, and does not include effects such as magnetic fields, externally driven turbulence, and cosmic rays.

The two terms on the right hand side of equation \eqref{eq:diff_eq2}, the momentum equation, represent the gravitational and hydrodynamic drag forces respectively. In the fixed mass case, these two terms balance one another in steady-state, and the corresponding velocity is the hydrodynamic drag terminal velocity. However, if $\Dot{m} > 0$, which is the case when the cloud is growing via mass accretion, then the cloud will slow down by momentum conservation. 
It can be shown that in these shearing cloud systems, the cloud develop a long tail-like structure and the hydrodynamic drag term is hence small compared to  $\Dot{m}v$ (see \citet{tan23} for a detailed argument). The terminal velocity in this case becomes $v_{\rm T,grow} = g t_{\rm grow}$, i.e. the acceleration of the cloud under gravity is predominantly balanced by the momentum transfer from background accretion instead of hydrodynamic drag forces. $\Dot{m}v$ is termed as `accretion drag'.

The crucial term in this model is thus the growth timescale $t_{\rm grow}$, which is determined by the physics of the turbulent radiative mixing layers that govern the rate of exchange of mass, momentum, and energy between the cloud and the background medium. 
This is the key refinement over earlier models. As mentioned above, in the absence of cooling, clouds are destroyed by hydrodynamic instabilities on the `cloud crushing' timescale $t_{\rm cc}$. In wind tunnels, this timescale is evaluated using the initial shear between the wind and the cloud, where the difference is maximal. Survival then depends on the ratio of $t_{\rm cc}$ and the cooling time. However, for a cloud infalling under gravity, two changes have to be made. Firstly, $t_{\rm cc}$ cannot be evaluated initially since the shear is originally zero and the cloud initially accelerates. $t_{\rm cc}$ thus depends on the instantaneous velocity or, if it exists, the terminal velocity. Second, since the cloud does not decelerate, the relevant growth timescale is not just the cooling time but $t_{\rm grow}$. This turns out to be a more stringent requirement, but we refer to the original paper for a full discussion. Assuming that infalling clouds meet this survival requirement, we can formulate a quantitative description of $t_{\rm grow}$ as follows.

The cloud mass growth rate can be decomposed as:
\begin{align}
    \Dot{m} \sim \rho_{\rm hot} A_{\rm cloud} v_{\rm mix},
    \label{eq:mdot}
\end{align}
where $\rho_{\rm hot}$ is the background density, $A_{\rm cloud}$ is the effective surface area of the cloud (note this is not the same as the cross-sectional area $A_{\rm cross}$) and $v_{\rm mix}$ is the velocity corresponding to the mass flux of the background onto this surface. If we write $m \sim \rho_{\rm cold} A_{\rm cloud} r$, this gives: 
\begin{equation} 
    t_{\rm grow} \sim \chi \frac{r}{v_{\rm mix}}. 
    \label{eq:tgrow-simple}
\end{equation}
Since $v_{\rm mix}$ is determined by the physics of radiative turbulent mixing layers, we can use the expression of $v_{\rm mix}$ from plane parallel simulations \citep{tan21}: 
\begin{equation}
    v_{\rm mix} \sim u^{\prime 3/4} \left( \frac{r}{t_{\rm cool}} \right)^{1/4}
    \sim v_{\rm shear}^{3/5} v_0^{3/20} \left( \frac{r}{t_{\rm cool}} \right)^{1/4}
    \label{eq:vmix-simple}
\end{equation}
where $t_{\rm cool}$ is the {\it minimum} cooling time in the mixing layer and $u^{\prime}$ is the peak turbulent velocity in the mixing layer (usually corresponding to intermediate temperature gas). Setting $v_{\rm shear} \sim v_{\rm T,grow} \sim g t_{\rm grow}$ yields: 
\begin{equation} 
    t_{\rm grow} \sim \chi^{5/8} \frac{r^{15/32} t_{\rm cool}^{5/32}}{g^{3/8} v_0^{3/32}}; \ \ {v_{\rm T,grow} < c_{\rm s,hot}}.
    \label{eq:tgrow_nonumbers}
\end{equation}
In the derivation above, we assume that the velocities remain in the subsonic/transonic regime. However, sufficiently large clouds can reach velocities exceeding the hot speed sound gas, in which case the scalings we have use saturate as the sound speed \citep{yang22}. In this case, from equations \eqref{eq:tgrow-simple} and \eqref{eq:vmix-simple}, we obtain:
\begin{equation}
    t_{\rm grow} \propto \frac{\chi  r^{3/4} }{{\rm c_{\rm s,hot}^{3/5}} t_{\rm cool}^{1/4}}. 
\label{eq:tgrow-sat-simple}
\end{equation}
The scaling relations presented can be further calibrated to the radiative turbulent mixing layer simulations. For cooling dominated regimes, \citet{tan21} find that $v_{\rm mix}$ follows
\begin{equation}
    v_{\rm mix} \approx 
    9.5 {\mathrm{~km}\,\mathrm{s}^{-1}}
    \left(\frac{u^{\prime}}{50 \, {\rm km \, s^{-1}}}\right)^{3/4} 
    \left(\frac{L_{\rm turb}}{100\,{\rm pc}}\right)^{1/4}
    \left(\frac{t_{\rm cool}}{0.03\,{\rm Myr}}\right)^{-1/4},
    \label{eq:vmix}
\end{equation}
where $L_{\rm turb}$ is the outer scale of the turbulence. For shearing layers, $u'$ is given by
\begin{equation}
    u' \approx 50{\mathrm{~km}\,\mathrm{s}^{-1}}
    \mathcal{M}^{4/5}
    \left(\frac{c_{\rm s,hot}}{150 {\mathrm{~km}\,\mathrm{s}^{-1}}} \right)^{4/5}
    \left(\frac{t_{\rm cool}}{0.03\,{\rm Myr}}\right)^{-0.1} ,
    \label{eq:uprime} 
\end{equation}
for $\chi\gtrsim 100$ and $\mathcal{M}\equiv v_{\rm shear} / c_{\rm s,hot}$.

One key assumption is equations \eqref{eq:vmix} and \eqref{eq:uprime} is that turbulence is already fully developed. This is not the case at the beginning since the cloud begins at rest. There is a transient period during which turbulence develops around the cloud via shear instabilities as it falls. The model hence includes a time dependent weight factor $w_{\rm kh}(t)$ that accounts for the initial onset phase of turbulence by modifying the mixing velocity, such that ${v}_{\rm mix} \rightarrow w_{\rm kh}(t) v_{\rm mix}$. In the absence of a more detailed physical model, \citet{tan23} used a simple ansatz for $w_{\rm kh}$ based on the the KH timescale $t_{\rm kh} \propto t_{\rm cc} \equiv \sqrt{\chi} r/v$ \citep{klein94},
\begin{equation} 
    w_{\rm kh}(t) = \min\left(1, \frac{t}{f_{\rm kh} t_{\rm cc}}\right)
\end{equation} 
This represents $v_{\rm in}$ growing linearly in time over an instability growth timescale which is proportional to the `cloud crushing' time, so $t_{\rm kh} = f_{\rm kh} t_{\rm cc}$ with a constant of proportionality $f_{\rm kh}$, until turbulence is fully developed where it caps at unity. We can rewrite this using $t/t_{\rm cc} = vt/(\sqrt{\chi}r)$ and noting that for a cloud initially at rest falling with constant acceleration, $z \propto vt$ so:
\begin{equation} 
    w_{\rm kh}(t) = 0.02 \frac{r}{z}.
\label{eq:w_kh} 
\end{equation} 
where we use $\chi = 100$ and the constant is calibrated from simulations in \citet{tan23}. The assumption of constant acceleration is valid at early times when the cloud is falling ballistically, before accretion drag becomes important.

The cloud surface area $A_{\rm cloud}$ follows a scaling relation $A_{\rm cloud} \propto (m/\rho_{\rm cloud})^{5/6}$ which accounts for both the mass growth and the formation of a lengthening tail during infall \citep{tan23}.
The cloud surface area is thus
\begin{align}
    A_{\rm cloud} \approx A_{\rm cloud,0} \left( \frac{m}{m_0} \frac{\rho_{\rm cloud,0}}{\rho_{\rm cloud}} \right)^{\alpha},
    \label{eq:acloud}
\end{align}
where $A_{\rm cloud,0}$, $\rho_{\rm cloud,0}$ and $m_0$ are the initial cloud surface area, density and mass respectively. The cloud density $\rho_{\rm cloud}$ changes because the ambient pressure increases as the cloud falls in a stratified medium, compressing the cloud. From equations~\eqref{eq:mdot}, \eqref{eq:vmix}, \eqref{eq:w_kh} and \eqref{eq:acloud}, we can write the growth time as
\begin{align}
    t_{\rm grow} = \frac{t_{\rm grow,0}}{w_{\rm kh}(t)}
                   \left( \frac{c_{\rm s,150}}{v} \right)^{3/5}
                   \left( \frac{t_{\rm cool}}{t_{\rm cool,0}} \right)^{1/4}
                   \left( \frac{m}{m_0} \frac{\rho_{\rm hot,0}}{\rho_{\rm hot}} \right)^{1-\alpha},
    \label{eq:tgrow_scaling}
\end{align}
where $c_{\rm s,150} = 150\,{\rm km\,s}^{-1}$ is the sound speed of gas at $10^6$~K and the initial growth time $t_{\rm grow,0}$ is given by
\begin{align}
    t_{\rm grow,0} \approx 35\,{\rm Myr }\,
                     \left( \frac{f_{\rm A}}{0.23} \right)
                     \left( \frac{\chi}{100} \right)
                     \left( \frac{r}{r_{100}} \right)
                     \left( \frac{L_{\rm turb}}{L_{100}} \right)^{-1/4}
                     \left( \frac{t_{\rm cool,0}}{0.03\,{\rm Myr}} \right)^{1/4}.
    \label{eq:tgrow_scaling2}
\end{align}
where $r_{100} = L_{100} = 100$\,pc and $r$ is the initial cloud size. We assume generally that $L_{\rm turb} \sim r$ since the hydrodynamic instabilities which drive turbulence have an outer scale set by the cloud size. $f_{\rm A}$ is a normalization factor to account for uncertainties arising from geometrical differences between the single mixing layers in \citet{tan21} and infalling clouds. From simulations in \citet{tan23}, $f_{\rm A} \sim 0.23$.

This model does not include a range of physical processes such as magnetic fields and cosmic rays whose importance and impact in these systems remain open questions and active avenues of research \citep{gronnow22,huang22,armillotta22}. We also do not include explicit viscosity or thermal conduction (although mass transfer is dominated mostly by turbulent diffusion, \citealp{tan21}). Lastly, while negligible to the dynamics for most of our parameter range of interest, self-gravity has been found to matter in the more compact HVCS \citep{sander21}.

\subsection{Simulations}
We carry out our simulations using the publicly available magnetohydrodynamical (MHD) code Athena\verb!++! \citep{stone20}. All simulations are run in 3D on regular Cartesian grids using the HLLC approximate Riemann solver and Piecewise Linear Method (PLM) applied to primitive variables for second order spatial reconstruction. By default, we use the third-order accurate Runge-Kutta method for the time integrator. 

Our simulation setup consists of a cloud initially at rest allowed to fall freely under gravity. We assume a constant gravitational acceleration $\boldsymbol{g} \equiv - g \hat{z} $, with $g = 10^{-8}$\,cm\,s$^{-2}$, taken from the fit in \citet{benjamin&daly97} for distances between 1 and 10\,kpc for the Milky Way. The simulation box has dimensions $10^2 \times 80$\,$r_{\rm cloud}$, with a resolution of $256^2\times2048$, meaning we resolve $r_{\rm cloud}$ by $\sim 25$ cells. As the cloud falls, we continuously adjust the frame of reference to match that of the center of mass of the cold gas in the box, where cold is defined to be gas below a temperature of $T \sim 2 \times 10^4$\,K. The cloud falls through a background of static hot $T_{\rm hot} = 10^6$\,K gas of density $n_0 = 10^{-4}\ccm$. This constant background is unphysical since there are no pressure gradients in the background to counteract gravity, meaning it is not in hydrostatic equilibrium. While \citet{tan23} explores more realistic background profiles, the simulations in this paper are restricted to this more idealized setup that allows us to study the physical mechanism without the confounding effects of a varying stratified background. This allows us to perform a cleaner test of the analytic model. Hence, to prevent the background from falling under gravity, we also introduce an artificial balancing force $\rho_{\rm hot}g$ upwards. The hot background thus feels a net zero force from gravity, while the cold cloud is negligibly affected. The initial density of the cloud is randomly perturbed at the percent level throughout the cloud to avoid numerical artifacts from initial symmetries. Boundary conditions are outflowing except at the bottom of the box where the background profile is enforced in the ghost cells are the velocity is set to match the frame velocity. We assume collisional ionization equilibrium (CIE) and solar metallicity ($X=0.7$, $Y=0.28$, $Z=0.02$) for radiative cooling. Our cooling curve comes from a piece-wise power law fit to the cooling table given in \cite{Gnat2007} over 40 logarithmically spaced temperature bins, starting from a temperature floor of $10^4$\,K, which we also enforce in the simulation. Our cooling implementation follows the fast and robust exact cooling algorithm described in \citet{townsend}. To emulate the effect of heating and to prevent the background medium from cooling over simulation timescales, we cut off any cooling above $5 \times 10^5$\,K. The particular choice of this value is unimportant \citep{gronke18,gronke20-cloud,abruzzo21}.

Our data suite consists of 13 simulations, 7 of which are used for training and 6 for testing. We use the same initial conditions for all simulations, with the only difference being the initial cloud size. The cloud sizes range from 30 pc to 1 kpc. Varying the cloud size varies the growth rates of the clouds and hence their mass growth and velocity histories.
Each cloud size generates up to 1000 data points spanning a time range of up to 225 Myr. Each data point consists of a center of mass velocity (of cold gas), total mass (of cold gas), and total distance fallen (with respect to center of mass) by the cloud. This corresponds to $v,m,z$ in the model.
This is a relatively small dataset, since running the large high resolution 3D simulations is computationally expensive. However, we will show that we are still able to achieve good results in this context.

\subsection{Neural ODEs and Symbolic Regression}
In this paper, we use neural networks embedded in the ODEs that represent the model (equations \eqref{eq:diff_eq1} -- \eqref{eq:diff_eq3}) to replace the weight factor $w_{\rm kh}$ (instead of using the form of the original ansatz) and later on, the dimensionless drag coefficient $C_0$ (hence allowing it to potentially vary in time as opposed to being an arbitrary constant). We then apply Symbolic Regression on the neural networks to generate the equations governing the terms, with the aim being equation discovery. In effect, this converts the neural networks into the form of symbolic equations. We apply this process to two sets of data. The first is a mock dataset we generate directly by evaluating the original model with the original analytical ansatz used for $w_{\rm kh}$. This is meant to test the effectiveness of the methodology when the underlying generating model (the `ground truth') is known to us. The second uses the suite of full 3D simulations described above.

The neural networks consist of a multilayer perceptron (MLP) with 3 hidden layers of 32 nodes each. Each layer uses a GELU activation function. Weights are initialized using a Xavier/Glorot initialization scheme. We use the \verb!ADAMW! optimizer, and train the network in 2 stages. In the first stage, we train for 1300 steps with a learning rate of $2 \times 10^{-3}$ on the first 20\% of the time series. In the second stage, we train for 3500 steps with a learning rate of $3 \times 10^{-3}$ on the entire training data. This is a common strategy in training neural ODEs to avoid getting initially trapped in local minima. The neural networks takes in $(z,v, m)$ and the initial cloud size $r$ as input. All input data into the neural network are normalized, and the mass which can grow by orders of magnitude over the simulation is log-transformed. The output of the network is a single value corresponding to $w_{\rm kh}$ or $C_0$.

For the neural network corresponding to $w_{\rm kh}$, we calculate our loss function as the mean squared error (MSE) between the predicted and actual mass solutions. We find that only including the mass $m$ (velocity $v$) in the loss function instead of all changing variables $(z,v$ and $m)$ improves training stability since the mass is the quantity most directly sensitive to the growth rate. We note that since the equations are coupled, the network is still able to make predictions for all variables. We also multiply the loss by a coefficient which decreases over time. This coefficient starts at 1 and then decreases linearly over time to a low of 0.1. This focuses the model on earlier times of the solution, which improves the training process in neural ODEs where late time values depend on earlier ones. For the second neural network corresponding to $C_0$, we calculate our loss function as the mean squared error (MSE) between the predicted and actual velocity solutions instead. This is because the drag term only directly affects velocity. Since we start with the first neural network above already trained and frozen in, we train on the full time evolutions from the start. We also `initialize' the output to be the value of unity (the pre-trained value) by simply adding 1 to the predicted value of $C_0$ (which is initially 0). We employ early stopping to avoid overfitting, and end the training when the loss on the test set stops improving. We thus train over 14 steps (1400 epochs).
We numerically solve the differential equation using Tsitouras' 5/4 method \citep{tsitouras2011runge}, which is a 4th order explicit Runge-Kutta method when using adaptive step sizing, which we employ via a Proportional-Integral-Derivative (PID) controller \citep{soderlind2003digital}. To construct, solve, and train the neural ODE, we use the \verb!JAX! framework \citep{jax2018github} heavily, along with the software libraries built on top of it, \verb!Diffrax! \citep{kidger22} for differential equations and \verb!Equinox! \citep{kidger2021equinox} for neural networks. This enables us to train on a single CPU core on the order of minutes.

For Symbolic Regression, we use the package \verb!PySR! \citep{cranmerInterpretableMachineLearning2023}. We include the physical units of the quantities when running Symbolic Regression --- this steers the discovery process towards dimensionally consistent equations. At the end, the model with the highest score is selected (the score is defined to be the negated derivative of the log-loss with respect to complexity). We allow the common binary operators along with the power and minimum operators, and the logarithm and exponentiation unary operators. We use a population size of 100 and ensure that we iterate for a sufficient amount of time to get a converged model.

\section{Results} \label{sect:results}

\subsection{Performance on Mock Data}
\begin{figure}
   \centering
   \includegraphics[width=\columnwidth]{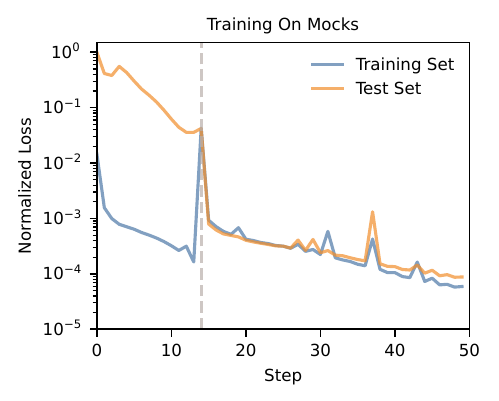}
   \caption{The learning curve (blue) for training the neural network along with the validation loss (orange) on the test set at 49 steps. Each step corresponds to 100 epochs of training. The transition between the two training stages occurs at step 15, indicated by a vertical dashed line. The loss saturates by the end of the training.}
   \label{fig:1}
\end{figure}
\begin{figure*}
   \centering
   \includegraphics[width=\textwidth]{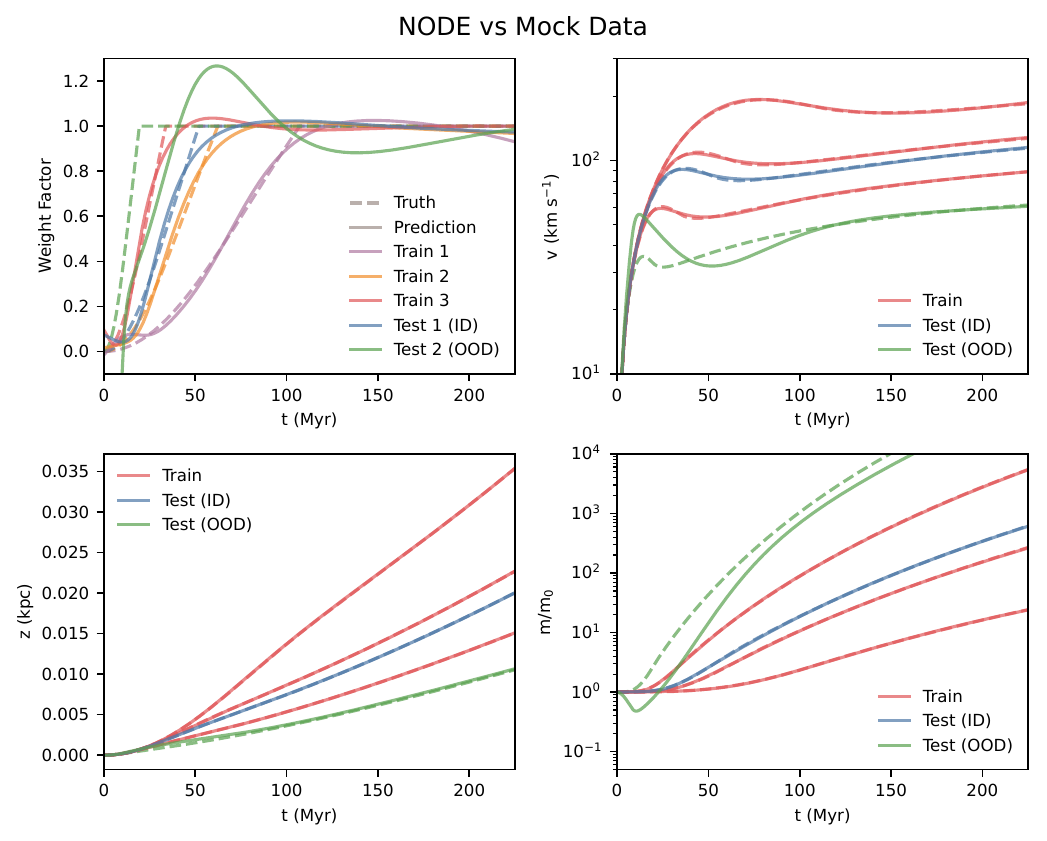}
   \caption{Comparison of neural ODE predictions against mock data. The four panels show (clockwise from top left): (a) Weight factor evolution, (b) Velocity evolution, (c) Mass evolution, and (d) Distance fallen. Mass is normalized by the initial mass $m_0$ at $t=0$. Dashed lines represent the ground truth, while solid lines represent the neural ODE predictions. Each panel compares the two over a range of cloud sizes from within the training data and for test cloud sizes both in-distribution (ID) and outside-of-distribution (OOD) of the training data.}
   \label{fig:2}
\end{figure*}
\begin{figure*}
   \centering
   \includegraphics[scale=0.9]{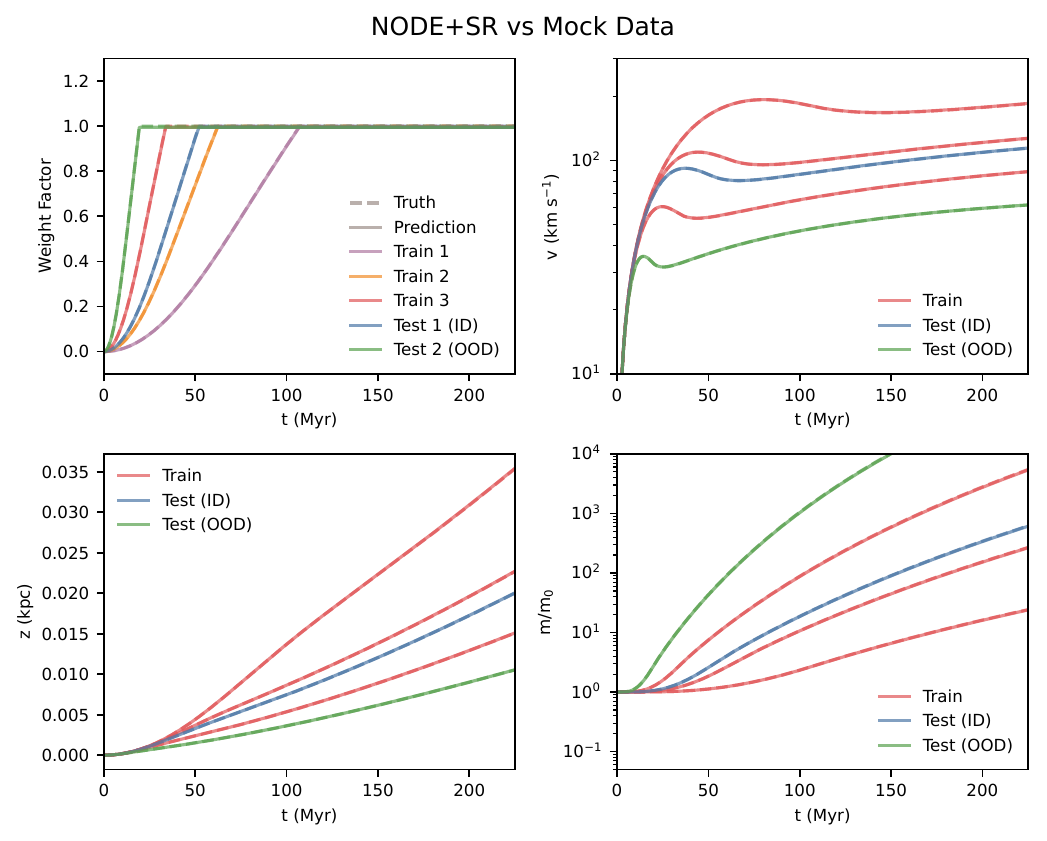}
   \caption{Same as Figure~\ref{fig:2}, but after applying Symbolic Regression to the neural ODE model. The SR model improves the predictive power of the neural ODE model, especially for out-of-distribution (OOD) test data. In this case SR recovers the ground truth equation used to generate the mock data.}
   \label{fig:3}
\end{figure*}
\begin{figure}
   \centering
   \includegraphics[width=\columnwidth]{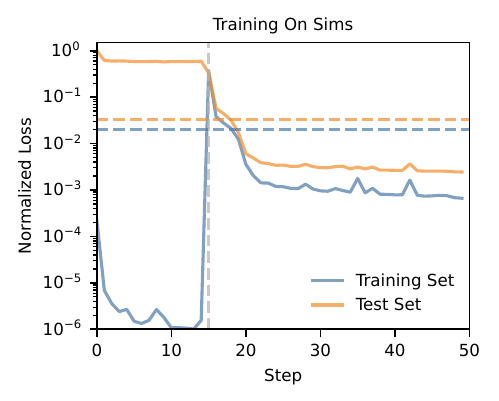}
   \caption{As in Figure~\ref{fig:1}, but for training the neural ODE model on actual simulation data. Horizontal dashed lines show the evaluation of the loss function between the simulation data and the model used in \citet{tan23} (Equation \eqref{eq:w_kh}).}
   \label{fig:4}
\end{figure}
\begin{figure*}
   \centering
   \includegraphics[width=\textwidth]{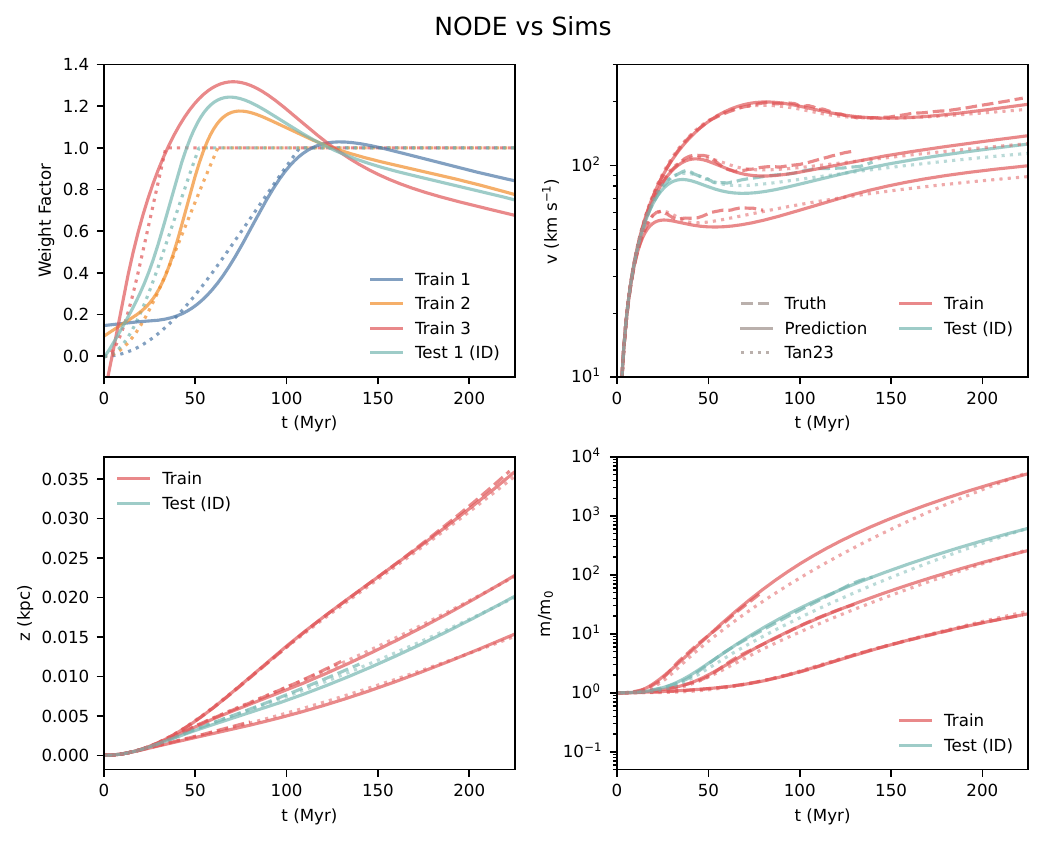}
   \caption{Comparison of neural ODE predictions (solid) against simulation data (dashed) and the \citet{tan23} model (dotted). Note that the `ground truth' for the weight factor in the simulations is unknown\protect\footnotemark[1]. Dashed lines show the more limited data from the simulations versus the mocks. The neural ODE model is able to capture the evolution of the cloud well, but less so for the velocity. This is like due to model not having a good treatment of hydrodynamic drag, in particular for the smaller clouds, a factor which only affects the velocity directly.}
   \label{fig:5}
\end{figure*}
\begin{figure*}
   \centering
   \includegraphics[width=\textwidth]{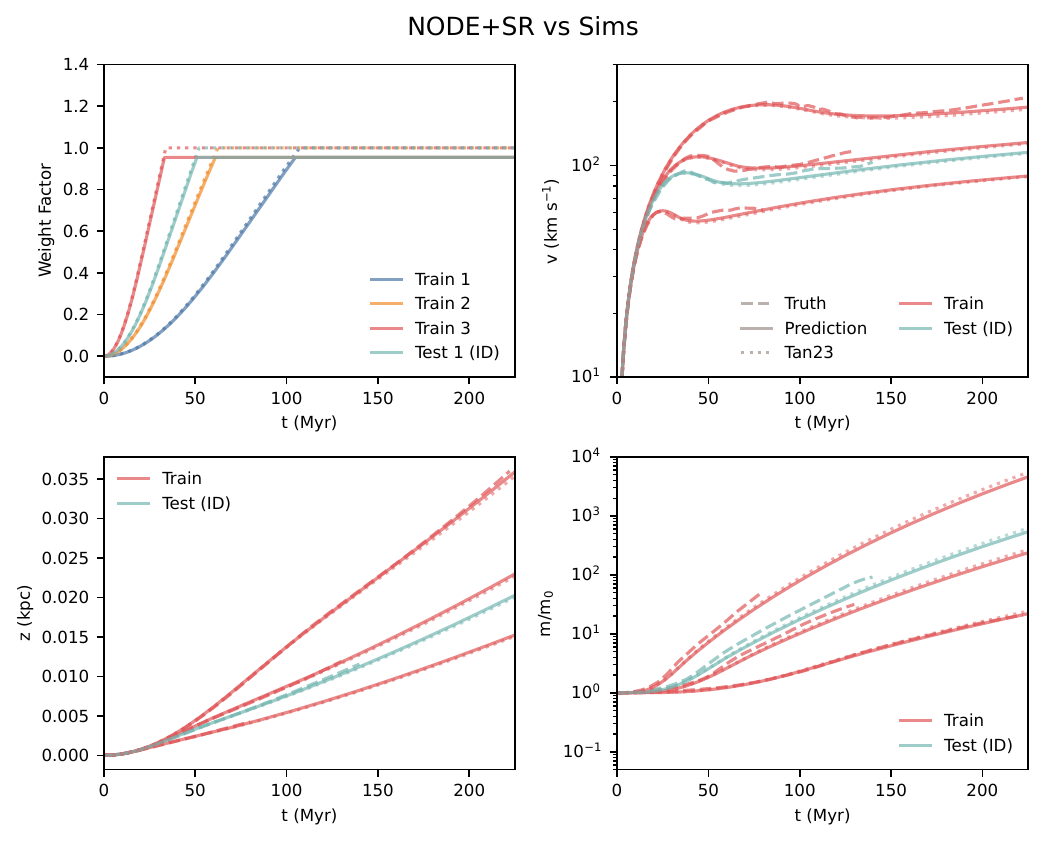}
   \caption{Same as Figure~\ref{fig:5}, but again after applying Symbolic Regression to the neural ODE model. Surprisingly, we recover a very similar equation to Equation \eqref{eq:w_kh}, indicating that the original guess was a good one.}
   \label{fig:6}
\end{figure*}

We first train on the mock data we generate to evaluate the performance of the neural ODE on a known hidden function $w_{\rm kh}$. As mentioned, we generate this dataset directly by evaluating the original model with the original analytical ansatz used for $w_{\rm kh}$ (Equation \eqref{eq:w_kh}). In Figure~\ref{fig:1}, we show the learning curve (in blue) for the training of the neural network, along with the validation loss (orange) on the test set at 49 steps. Each step corresponds to 100 epochs of training. The transition between the two training stages is indicated by a vertical dashed line. The initial loss of the training set is small since we only train on 20\% of the full time series. The loss with respect to the training and test set are similar once we reach the second stage, where the full time series is used during training. The validation loss on the test set eventually saturates by the end of the training process, while the loss on the training set continues to decrease as the neural network begins to overfit to the training data.

In Figure~\ref{fig:2}, we show the comparison of the trained neural ODE predictions against the mock data. The four panels show the evolution of (clockwise from top left): the weight factor, the velocity, the cloud mass, and the total distance fallen. Cloud mass is normalized by the initial mass $m_0$ at $t=0$. Dashed lines represent the ground truth generated by the assumed model, while solid lines represent the neural ODE predictions. Each panel compares the two over a range of cloud sizes from within the training data (red lines) and for test cloud sizes both in-distribution (ID, teal line) and outside-of-distribution (OOD, green line) of the training data. 
Cloud sizes are considered ID if they lie within the smallest and largest cloud of the training set, and OOD if they lie outside. The training data consists of cloud sizes ranging from 30~pc to 1~kpc, of which (30~pc, 100~pc and 300~pc) are displayed in Figure~\ref{fig:2}. The ID test data consists of cloud sizes ranging from 40~pc to 800~pc, of which 70~pc is displayed in Figure~\ref{fig:2}. The OOD cloud in Figure~\ref{fig:2} has a size of 10~pc.
The neural ODE is able to reproduce the velocity, mass, and distance evolution of the training data well, and similarly so for test clouds with sizes that fall within the range of sizes spanned by the training data. 
We can quantify this by computing average percentage error of the model prediction relative to the ground truth. For the normalized mass evolution histories, the mean percentage error range from 0.6\% - 0.95\% for the training curves, and is 1\% for the ID curve.
We can see that this is directly reflected in the ability of the neural network to predict the weight factor. 
The mean percentage error of the weight factor (ignoring the first 20 Myr since the weight factor is close to zero there) for the training curves range from 1.5\% - 3.5\%, and is 2.5\% for the ID curve.
However, the neural ODE does a poorer job at predicting the weight factor and hence the mass and velocity evolution for the test cloud with a size that is just outside the range of the training data. 
For the OOD curve, the mean percentage error for the weight factor is 11\% and 44\% for the mass evolution.
This is not surprising --- generalizing beyond the parameter space of the training data is a well-known challenge in scientific machine learning. Figure~\ref{fig:2} also shows that the mass evolution is the most sensitive to the weight factor, since the initial error in mass prediction grows over time (the log scaling of the vertical axis in the mass and velocity plots underplay this difference).

Understanding the physics of the system is also an important goal. To this end, we explore the application of Symbolic Regression to the neural ODE model. Since we are using a neural network in our ODE to model the influence of a specific process (in this context the development of turbulence in the mixing layers), we have defined the scope of the physical process the neural network is capturing. By doing this, it should be possible to use Symbolic Regression to discover the equation governing this process since we do not expect the underlying equation to be extremely complex. Of course, there is an implicit assumption here that the rest of the ODE is a good reflection of the physics of the rest of the system. If this is not the case, then the neural network might `learn' things related to other parts of the system, which would muddy the interpretation.

In Figure~\ref{fig:3}, we show the results of applying Symbolic Regression to the neural ODE model. This involves replace the neural network with the found symbolic representation and solving the resulting set of equations. In this case, SR `discovers' the ground truth equation used to generate the mock data (Equation \eqref{eq:w_kh}). Table~\ref{tab:phases} shows a comparison between the two equations, which are almost identical. 
As before, we compute the mean percentage error for the weight factor and the mass evolution. For the weight factor, all mean percentage errors are now around 0.5\%, including the OOD curve. For the mass evolution, the percentage errors are between 0.4\% and 1.1\% for the training curves, 0.8\% for the ID curve, and 1.5\% for the OOD curve. The SR model thus improves the generalization of the neural ODE. It also makes the results interpretable --- we are able to interpret the discovered equation as with any other physical equation we might have derived using traditional methods.

\subsection{Performance on Simulation Data}

We now turn our attention to training the neural ODE model on actual simulation data. In Figure~\ref{fig:4}, we show the learning curve for the training of the neural network on the simulation data. The training is less smooth compared to with the mock data. The loss is much higher in the test set than the training set. This likely indicates that there is more overfitting to the training data. The improvement in the loss also starts to saturate at a much earlier time. The horizontal dashed lines show the evaluation of the loss function between the simulation data and the model used in \citet{tan23} (Equation \eqref{eq:w_kh}) for both the training and the test set, which represents the human benchmark for comparison. The lower losses indicate that the neural ODE model is better able to capture the evolution of the weight factor (or equivalently the mass growth rate) in the simulations.

In Figure~\ref{fig:5}, we show the comparison of the trained neural ODE predictions against the simulation data. Simulation data is represented by the dashed lines. It can be seen that the data from the simulation is more limited for smaller clouds due to box sizes in the simulation.
Since the box sizes scale with cloud sizes, cold gas starts interacting with the boundaries earlier for smaller clouds, leading to earlier termination of the simulations. One advantage of neural ODEs is being able to handle such inhomogeneities in the data easily.
The neural ODE model is able to capture the evolution of the mass well (lower right panel), but less so for the velocity (upper right panel). The average percentage error in mass and velocity is $4.4\%$ and $9.9\%$ for the worst training curve in Figure~\ref{fig:5}, and $3.9\%$ and $12.4\%$ for the test curve. We attribute this to our choice to use the mass evolution curves in the loss function. The results indicate that while the model is getting the mass growth right, it is likely not accounting for hydrodynamic drag well, which only directly affects velocity. This effect appears more pronounced for smaller clouds.
Again, we are able to draw the above conclusions due to the constrained scope of the neural network in our ODE. So although the predicted mass matches the simulation data well, the velocity profiles shows more deviation. This itself is useful information, as it indicates that we need to better model the drag in order to increase the fidelity of the model.

For comparison, we show the predictions of the \citet{tan23} model overlaid as dotted lines. While Figure~\ref{fig:5} shows our NODE model does better in matching mass growth over the original model, it does worse in matching the velocity evolution. This is expected since that model was hand calibrated to minimize errors in both mass and velocity. We will show in the next section that we can further improve our model by adding a second neural network to model the drag coefficient.

\begin{table}
    \centering
        \begin{tabular}{c|c|c}
            \hline 
            \hline
            Dataset & True Equation & Learned Equation (SR) \\
            \hline
            Model &  $\min\left(1, 0.02\frac{z}{r}\right)$  & $\min\left(0.9948299, 0.01994161\frac{z}{r}\right)$\\
            Simulation & - & $\min\left(0.95419395, 0.01955151\frac{z}{r}\right)$\\
            \hline
        \end{tabular}
    \caption{Symbolic regression learned equations selected by best score.}
    \label{tab:phases}
\end{table}

In Figure~\ref{fig:6}, we show the results of again applying Symbolic Regression to the neural ODE model. In this case, even when trained on simulation data, SR still recovers a very similar equation to Equation \eqref{eq:w_kh}, only with slightly different coefficients. The learned equation is shown in Table~\ref{tab:phases}. While slightly surprising, this suggests that this was a good choice in the original work despite its simplicity. However, the predicted mass evolution is less accurate.
The mean percentage error in mass and velocity for the worst training curve in Figure~\ref{fig:6} is 19\% and 4\%, and 21\% and 7\% for the test curve. Since we basically recover Equation \eqref{eq:w_kh}, we get back the predictions of the original \citet{tan23} model. In evaluating the fitness of equations during Symbolic Regression, complexity is penalized. It is likely that this equation captures the essence of the physics of the system, but is perhaps missing higher order effects. One possible avenue for future work is to adjust this complexity penalty further to explore equations which fit the data better but are more complex. These equations can then be analyzed for their physical relevance. 

\footnotetext[1]{This could be estimated directly from the simulation in principle by comparing $t_{\rm grow}$ from the simulations versus the analytic form, but is noisy in practice. We instead evaluate the performance by the predictions for the time evolution of mass and velocity.}

\subsection{Training a Second Neural Network}
\begin{figure}
   \centering
   \includegraphics[width=\columnwidth]{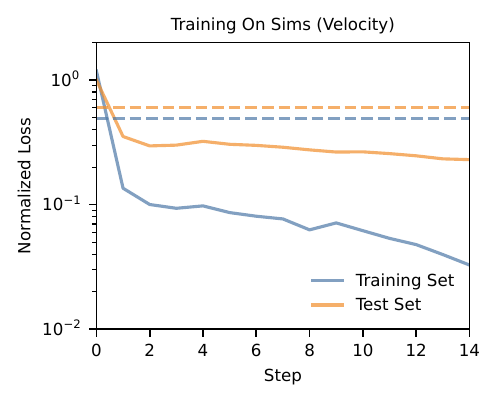}  
   \caption{The loss curves during training of the second neural network to learn the drag coefficient. As in Figure~\ref{fig:2}, horizontal dashed lines show the evaluation of the loss function between the simulation data and the model used in \citet{tan23} (Equation \eqref{eq:w_kh}).}
   \label{fig:7}
\end{figure}
\begin{figure*}
   \centering
   \includegraphics[width=\textwidth]{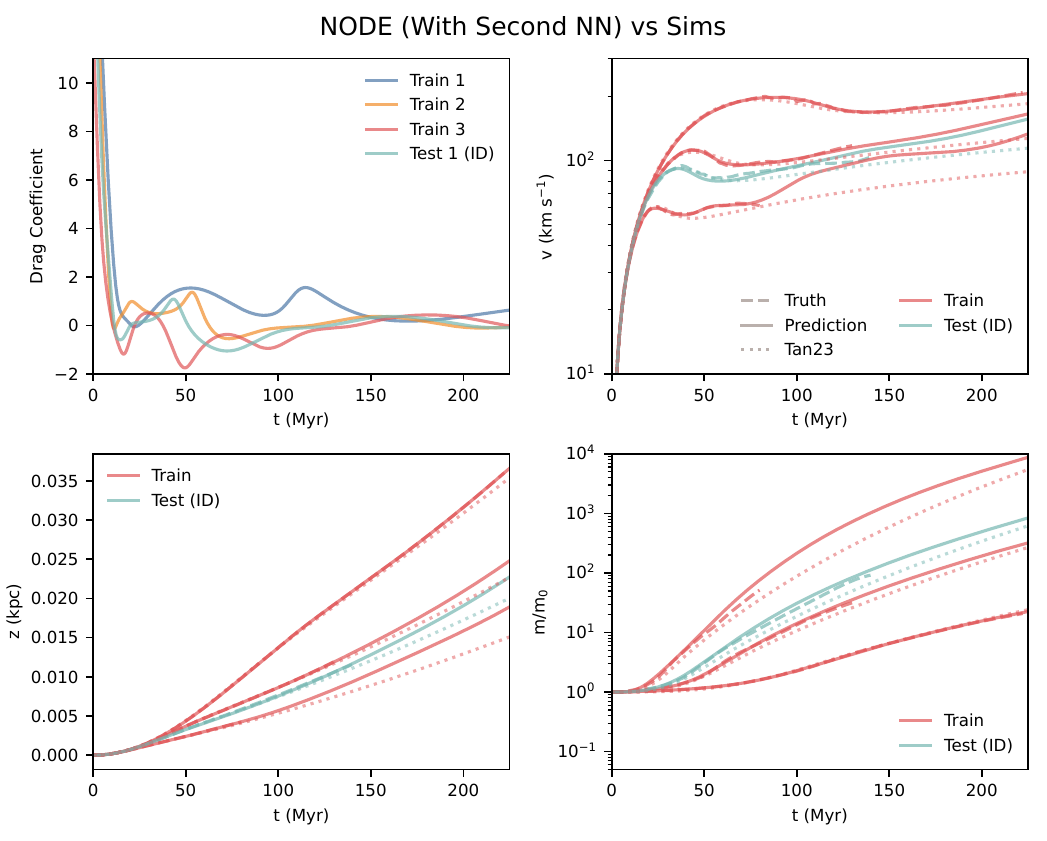}
   \caption{Results after training a second neural network to learn the drag coefficient over time, which is plotted in the top left panel. This quantity starts high and rapidly decreases to a small number. The total model now predicts the velocity evolution well. The mass predictions are only slightly less accurate.}
   \label{fig:8}
\end{figure*}

Up to this point, we have presented results for training a single neural network to learn the evolution of the weight factor describing the onset of turbulent mixing. However, as we noted above, our learned neural ODE does a good job in matching the mass evolution but differs from the simulation data in its prediction for the velocity evolution (see Figure~\ref{fig:5}). We attributed this to firstly, designing a loss function that only takes into consideration the mass evolution (physically motivated by the fact that turbulent mixing only directly affects mass growth), and secondly, the simplification of the representation of conventional drag in our equations. This is the rightmost term in Equation~\eqref{eq:diff_eq2}, where we assume that the drag coefficient $C_0$ is a constant equal to unity. In reality, the drag coefficient is affected by a range of factors that can influence the amount of drag, and can vary with parameters such as velocity and geometry. Due to the complexity involved, drag coefficients are usually measured experimentally in wind tunnel experiments. While this term is generally much smaller than accretion drag for growing clouds \citep{tan23}, we can further improve our subgrid model by repeating the process and replace $C_0$ with a second neural network.

We begin with the pre-trained model from the previous section with the learned weight factor and freeze the parameters of that first neural network. We then replace $C_0$, which was simply a constant of unity, with a new, second neural network which we then train. 
We adopt a similar loss function as before, but compute the loss with respect to the velocity instead of the mass, since $C_0$ only directly affects the clouds velocity. 
Figure~\ref{fig:7} shows the training and validation loss over the training process. 
As before, it seems the network overfits the training set. While the loss on the test set drops quickly initially, it is then followed by a longer period of slow improvement before saturating.
The results can be seen in Figure~\ref{fig:8}, which shows that the model does a better job at matching the velocity evolution (compared to Figure~\ref{fig:5}). The average percentage error in mass and velocity is $15.6\%$ and $0.87\%$ for the worst training curve in Figure~\ref{fig:5}, and $10.8\%$ and $4.85\%$ for the test curve. Although the average percentage error is now higher for the mass, the mass varies by two orders of magnitude over the course of the simulation, while the velocity varies only by $\sim10\%$ after an initial acceleration phase. The predictions for mass evolution have thus shifted since the equations are coupled, but only by a small margin since the velocity changes are small. We could continue to iterate the training of the neural networks to further reduce the errors, but the improvements are likely to be incremental. 
Note that training separately makes training more efficient since we were able to use two separate loss functions, with the first neural network capturing the more important physical uncertainty in the system.

Once again, we run symbolic regression on this second neural network to recover a physical expression for the drag coefficient for interpretability. This gives the expression
\begin{equation}
C_0 = \min (0.92/v, 8.6).
\end{equation}
The top left panel of Figure~\ref{fig:8} shows the predicted time evolution of the drag coefficient for various clouds as predicted by the neural network. They can be characterized by an initial rapid decline before saturating close to zero, which the learned equation captures. The minimum in the learned function is due to the function $1/v$ diverging at $t=0$ when the cloud is at rest, hence requiring a ceiling value to cap the fitness penalty. This inverse scaling with velocity is consistent with a range of empirical data for spheres which find that the drag coefficient scales inversely with the Reynolds number, which itself scales linearly with velocity \citep{goossens19}. Furthermore, while the cloud begins as spherical object, it develops a long tail and becomes more streamlined as it falls and grows. We would hence expect the drag coefficient to approach some small value $<1$, which is consistent with the predictions of the neural network and the learned equation. 

\section{Discussion and Conclusion} \label{sect:conclusions}
For systems where the full range of scales cannot be resolved, as is common in astronomy, we rely on subgrid models to capture important unresolved processes. A relevant example directly applicable to this work is subgrid modelling of cool clouds in cosmological simulations \citep{smith24}. Using theory to derive differential equation models and testing them against simulations is the traditional method of constructing these models. We have shown that (at least in the context of galactic cloud-environment interactions) neural ODEs together with Symbolic Regression can be used to improve the predictive power of such models and better capture the complex physics involved. For our system, the trained neural ODE model can accurately capture the evolution of the weight factor, velocity, mass, and distance fallen for both mock data and simulation data. If we consider the mean percentage errors of the mass evolution history of the test cloud in the figures above, the original model has an average error of 21\% for mass and 7\% for velocity. The trained model using just one neural network for the weight factor has an average error of 4\% for mass and 12\% for velocity, thus greatly reducing the error for the predicted mass evolution history. Symbolic regression on the first neural network recovers the original model. Training a second neural network (with the first frozen in) to account for non-constant drag further changes leads to a 11\% error for mass and 5\% for velocity. This is strictly better than the original model, and could likely be improved via iteration or parallel training of the two networks.
These improvements are in spite of the small size of the training data. By leveraging modern software libraries, constructing and training of the neural ODE model is seamless and rapid. By embedding the neural network in the ODEs, we are able to constrain its scope of impact on the system. This makes the neural ODE model both interpretable and physically motivated, while still being flexible and predictive. It also allows us to effectively employ symbolic regression for scientific equation discovery by constraining the complexity of the search space. Symbolic Regression can be used to improve the interpretability of the neural ODE model. For the mock data, SR recovers the ground truth equation used to generate the data. In the case of the simulation data, SR recovers a very similar equation to the one used in the original work for the weight factor, and provides insight into the time evolution of the drag coefficient not considered in the original model. We have also demonstrated that we can train multiple neural terms in the ODE, one for the weight factor and one for the drag coefficient, each with physically tailored loss functions (one using mass and the other velocity). This partitioning not only further improves the predictive power of the model (compared to just adjusting the weight factor for example), but also increases the effectiveness of using Symbolic Regression to make physical insights.

It is important to highlight that the simulations we have run in this work are extremely idealized. In a more realistic setting with less idealized initial conditions, the time-dependent weight factor might be more complicated, for example depending on turbulence inherent in the background seeded by other physical mechanisms such as extrinsic turbulence or cooling induced pulsations in the cloud \citep{gronke20-mist,gronke22}. A natural follow up would be to apply the methodologies explored in this work to these more realistic setups. 

Nonetheless, we have shown how neural ODEs in tandem with Symbolic Regression are natural extensions of traditional methods of formulating subgrid or semi-analytic models and would be valuable parts of the modelling toolkit, both to develop more accurate subgrid models of complex systems and to enhance the scientific discovery process.

\section*{Acknowledgements}
We thank Shirley Ho, Dustin Nguyen and Stephanie Tonnesen for insightful discussions, and the anonymous referee for a helpful and constructive report. Besides the packages mentioned in the text, we have also made use of the software packages \verb!matplotlib! \citep{Hunter2007} and \verb!numpy! \citep{van2011numpy} whose communities we thank for continued development and support. BT is supported by the Simons Foundation through the Flatiron Institute. 

\section*{Data Availability}
We make our code and all data used in this work public in a GitHub repository (\href{https://github.com/zunyibrt/Neural-Infalling-Cloud-Equations}{github.com/zunyibrt/Neural-Infalling-Cloud-Equations}).

\bibliography{references}
\bsp
\label{lastpage}
\end{document}